Harnessing Avidity: Quantifying Entropic and Energetic Effects of Linker Length and Rigidity Required for Multivalent Binding of Antibodies to HIV-1 Spikes

Tal Einav[a], Shahrzad Yazdi[b], Aaron Coey[c], Pamela J. Bjorkman[c,*], and Rob Phillips[a,c,d,e,*]

[a]Department of Physics, California Institute of Technology, Pasadena, CA 91125

[b]Department of Materials Science and Engineering, Massachusetts Institute of Technology, Cambridge, MA 02139

[c]Division of Biology and Biological Engineering, California Institute of Technology, Pasadena, CA 91125

[d]Department of Applied Physics, California Institute of Technology, Pasadena, CA 91125

[e]Lead contact

*Correspondance: bjorkman@caltech.edu (P.J.B.), phillips@pboc.caltech.edu (R.P.)

**Abstract**
Due to the low density of envelope (Env) spikes on the surface of HIV-1, neutralizing IgG antibodies rarely bind bivalently using both antigen-binding arms (Fabs) to crosslink between spikes (inter-spike crosslinking), instead resorting to weaker monovalent binding that is more sensitive to Env mutations. Synthetic antibodies designed to bivalently bind a single Env trimer (intra-spike crosslinking) were previously shown to exhibit increased neutralization potencies. In initial work, diFabs joined by varying lengths of rigid double-stranded DNA (dsDNA) were considered. Anticipating future experiments to improve synthetic antibodies, we investigate whether linkers with different rigidities could enhance diFab potency by modeling DNA-Fabs containing different combinations of rigid dsDNA and flexible single-stranded DNA (ssDNA) and characterizing their neutralization potential. Model predictions suggest that while a long flexible polymer may be capable of bivalent binding, it exhibits weak neutralization due to the large loss in entropic degrees of freedom when both Fabs are bound. In contrast, the strongest neutralization potencies are predicted to require a rigid linker that optimally spans the distance between two Fab binding sites on an Env trimer, and avidity can be further boosted by incorporating more Fabs into these constructs. These results inform the design of multivalent anti-HIV-1 therapeutics that utilize avidity effects to remain potent against HIV-1 in the face of the rapid mutation of Env spikes.

**Significance**
IgG antibodies utilize avidity to increase their apparent affinities through simultaneous binding of two antigen-binding Fabs – if one Fab dissociates from an antigen, the other Fab can remain attached, allowing rebinding. HIV-1 foils this strategy by having few, and highly-separated, Envelope spike targets for antibodies, forcing most IgGs to bind monovalently. Here we develop a statistical mechanics model of synthetic diFabs joined by DNA linkers of different lengths and flexibilities. This framework enables us to translate the energetic and entropic effects of the linker into the neutralization potency of a diFab. We demonstrate that the avidity of multivalent binding is enhanced by using rigid linkers or including additional Fabs capable of simultaneous binding, providing the means to quantitatively predict the potencies of other antibody designs.

**Introduction**

Despite decades of research since its discovery, Human Immunodeficiency Virus-1 (HIV-1) continues to threaten global public health (1). While there have been advances in our understanding of the mechanisms of infection and the development of preventative and therapeutic strategies, there remains no cure for HIV-1 infection. Antiretroviral therapy with small molecule drugs can control the progression of the virus, allowing those infected with HIV-1 to live longer and healthier lives, but the treatment includes detrimental side effects, and when discontinued or not taken as prescribed, leads to viral rebound to pre-treatment levels (2). A major factor confounding the development of a prophylactic vaccine is the rapid mutation of HIV-1 which leads to the emergence of many new strains, even within a single individual (3). Thus, most antibodies raised by the host immune system are strain-specific or neutralize only a subset of strains, leading to viral escape from host antibodies.

Recent interest has focused upon the isolation of broadly neutralizing IgG antibodies (bNAbs) from a subset of HIV-1–infected individuals (4). These antibodies bind to and block the functions of the HIV-1 envelope (Env) spike, the viral protein responsible for the fusion of HIV-1 to the host cell (5). The discovery and characterization of HIV-1 bNAbs has brought new impetus to the idea of passively delivering antibodies to protect against or treat HIV-1 infection. bNAbs can prevent and treat infection in animal models (6-12) and exhibited efficacy against HIV-1 in human trials (13-16). However, HIV-1 Env mutates to become resistant to any single bNAb, as even the most potent NAbs developed in an infected individual normally fail to neutralize autologous circulating viral strains (17-20). As a result, antibodies that develop during HIV-1 infection appear to be unable to control the virus in an infected individual.

We previously proposed that one mechanism by which HIV-1 evades antibodies more successfully than other viruses arises from the low surface density of Env spikes that can be targeted by neutralizing antibodies (21, 22). Compared to viruses such as influenza A, dengue, and hepatitis B, the density of Env spikes on the surface of HIV-1 is about two orders of magnitude smaller (22). For example, influenza A has ≈450 spikes per virion, whereas each HIV-1 virion incorporates only 7-30 Env spikes (average of 14) (22-26), even though both influenza A and HIV-1 are enveloped viruses with ≈120 nm diameters (Fig. 1A). The HIV-1 spikes are the machinery by which the virus binds its host receptor CD4 and coreceptor CCR5/CXCR4 to mediate the fusion of the host and viral membranes that allows its genome to enter target cells (5). As a consequence of its small number of spikes, HIV-1 infection of target cells is inefficient; the transmission probabilities for sexually-acquired HIV-1 infection range from 0.4 to 1.4% (27). However, the reduced infectivity of HIV-1 comes with a concomitant reduction in the ability of antibodies to control the virus, as the surface spikes serve as the only targets for neutralizing antibodies that can block infection of target cells (4).

The close spacing of spikes on typical viruses allows IgG antibodies to bind bivalently to neighboring spikes (inter-spike crosslinking) using both of their antigen-binding arms (Fabs). However, most spikes on HIV-1 virions are too far apart (typically over 20 nm separation) (22) to permit inter-spike crosslinking by IgGs whose antigen-binding sites are separated by ≤15 nm (28). While each homotrimeric HIV-1 spike includes three binding sites (epitopes) for an antibody, the architecture of HIV-1 Envs prohibits simultaneous binding of two Fabs within a single IgG to the same Env (intra-spike crosslinking) (29, 30). We suggested that predominantly monovalent binding by anti-HIV-1 antibodies expands the range of Env mutations permitting antibody evasion, since reagents capable of bivalent binding through inter- or intra-spike crosslinking would be less affected by Env mutations that reduce but do not abrogate binding and thus may be more potent across multiple strains of HIV-1 (21, 22). The hypothesis that HIV's low spike numbers and low densities contributes to the vulnerability of HIV-1 bNAbs to

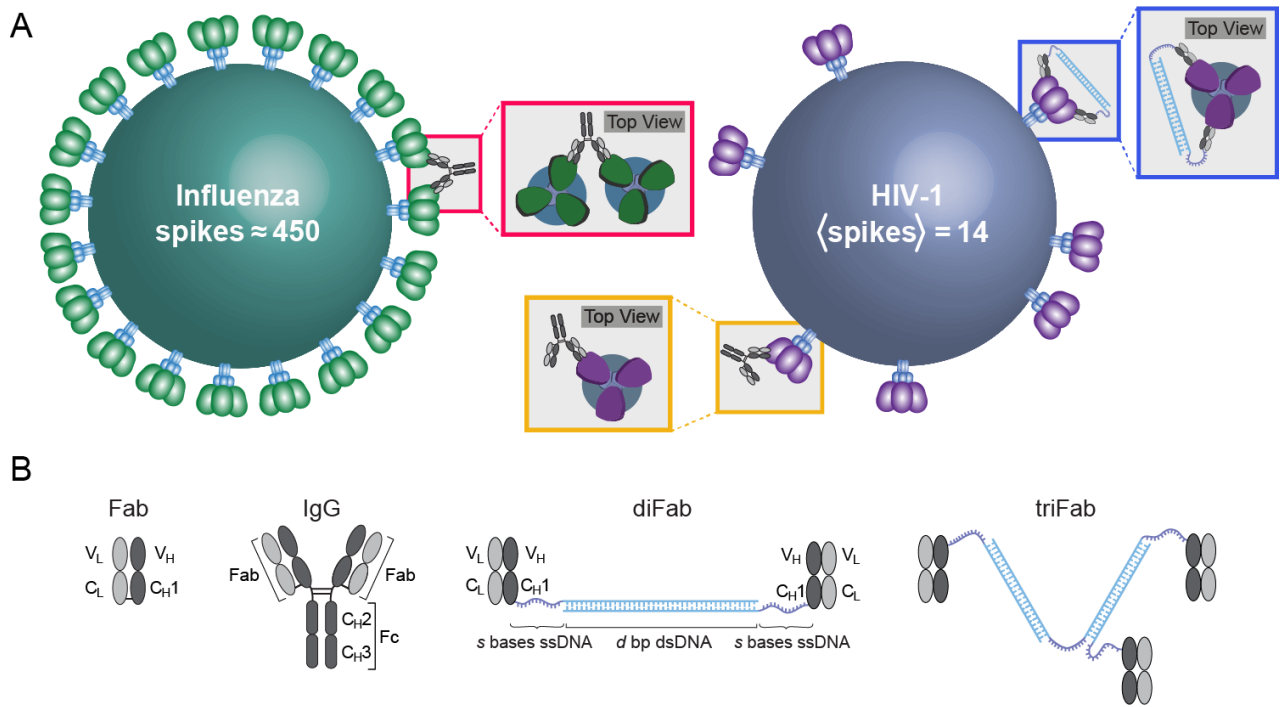

**Fig. 1. Effects of spike density on IgG binding.** (A) Close spacing of surface spikes on influenza A allows bivalent binding of IgGs to adjacent spikes (red boxes). In contrast, HIV-1 has few spikes (14 on average) spaced far apart, and because the HIV-1 spike architecture prohibits simultaneous binding of two Fabs to a single Env trimer, most IgGs bind monvalently to HIV-1 Envs (gold boxes). We investigated a synthetic diFab designed to bind bivalently to a single HIV-1 spike trimer (blue boxes). (B) Schematics of a Fab, an IgG, a diFab composed of two Fabs joined together by $d$ bp dsDNA and two segments of $s$ ssDNA bases, and a triFab made up of three Fabs.

---

spike mutations is supported by independent biochemical and EM studies demonstrating that HIV-1 has an unusually low number of spikes that are not clustered (23-26, 31), and that bivalent IgG forms of anti-HIV-1 NAbs are only modestly more effective than monovalent Fabs, by contrast to bivalent IgGs against other viruses, which can be 100s- to 1000s-fold more potent than counterpart monovalent Fabs (21, 22, 29, 30).

An antibody's neutralization potency against a virus is related to its antigen-binding affinity, which is defined as the binding strength between a Fab and its antigen (32) described by the equilibrium dissociation constant $K_D$ = [Fab][Ag]/[Fab–Ag], where [Fab], [Ag] and [Fab–Ag] are the concentrations of the antibody Fab, antigen, and the complex, respectively (33). In bivalent molecules interacting with binding partners that are tethered to a surface, the apparent affinity, or avidity, can be enhanced by multivalent binding. Such multivalent interactions are seen in many biological contexts including cell-cell communication, virus-host cell interactions, antibody-antigen interactions, and Fc receptor interactions with antigen-antibody complexes (34). Avidity effects benefit these interactions from both kinetic and thermodynamic standpoints. Binding bivalently to tethered binding partners is advantageous kinetically because if one arm dissociates, the likelihood of it finding its binding partner is greater due to the constraint of being tethered (35). Avidity effects are also advantageous thermodynamically; whereas binding the first arm results in losses of translational and rotational degrees of freedom, the subsequent binding of the second arm incurs a smaller entropic cost, thereby increasing the likelihood of the bivalent state (35).

In the context of an IgG with two antigen-binding Fabs, the ability to bind bivalently to a virus is dependent on geometric factors such as the separation distances and orientations of tethered epitopes either on adjacent spikes during inter-spike crosslinking (Fig. 1A, red box) or on individual spikes if intra-spike crosslinking can occur (Fig. 1A, blue box) (36). Because the large distances between HIV-1 spikes makes inter-spike crosslinking unlikely, in this work, we focus exclusively on the latter mechanism of achieving bivalent binding. Although IgGs are too small to intra-spike crosslink (Fig. 1A, gold box) (29, 30), we previously engineered larger reagents (homo- and hetero-diFabs) that were designed to bind to a single Env, resulting in mean neutralization potency increases over a panel of HIV-1 strains (21). These diFab constructs were composed of two IgG Fabs joined by different lengths of double-stranded DNA (dsDNA), which served as both a rigid linker and a molecular ruler to probe the conformations of HIV-1 Env on virions (21) (Fig. 1B). The dsDNA was flanked by two short single-stranded DNA (ssDNA) segments, where the primary differences between the two types of DNA is that dsDNA is more rigid and shorter ($0.34 \frac{\text{nm}}{\text{bp}}$) than the more flexible and longer ssDNA ($0.64 \frac{\text{nm}}{\text{base}}$).

In this work, we expand upon these earlier results and theoretically analyze whether changing the flexibility of the linker joining the two Fabs could also enhance neutralization potency. This enables us to compare a spectrum of possibilities from a rigid linker solely comprising dsDNA to a fully flexible linker composed of only ssDNA. To that end, we developed a statistical mechanical model to systematically evaluate the effects of linker length and rigidity on synergistic neutralization by a diFab. We then generalize our model to a triFab design and demonstrate that simultaneously binding to three Env epitopes can greatly boost avidity. Insights from our synthetic constructs can be adapted to antibody design in other systems in which length and rigidity of linkers in multivalent reagents must be balanced to elicit the most effective response.

## Results

**Estimating the Parameters of diFab Binding from Crystal Structures.** While HIV-1 Env fluctuates between multiple conformations, we assume that a diFab neutralizes the virus by binding to one specific state of Env at which the distance between the C-termini of the two Fabs (where the DNA is joined) is defined to be $l_{\text{linker}}$ (Fig. 2A). For example, the predicted distance based on modeling adjacent 3BNC60 Fabs on a low-resolution open structure of an HIV-1 trimer (37) was ≈20 nm. More recently, we used the 3BNC117 Fab-gp120 portion of a cryo-EM structure to measure the distance between adjacent Fab $C_H$1 C-termini in the closed conformation of Env and then modeled a 3BNC60-gp120 protomer into three recent cryo-EM structures of Env trimers in different conformations: an open Env bound by the b12 bNAb in which the coreceptor binding sites on the V3 loops are not exposed (38), an open CD4-bound Env structure with exposed V3 loops (38), and a partially-open CD4-bound Env in which the gp120 subunits adopted positions mid-way between closed and fully open (39). From these structures, we measured distances of 15.8, 20.3, 20.4, and 20.1 nm between C-termini of Fab $C_H$1 domains modeled onto the closed, b12-bound open, CD4-bound open, and CD4-bound partially-open Env conformations, respectively. Based on these values, we set $l_{\text{linker}}$=20 nm.

A further model parameter, $l_{\text{flex}}$=1 nm, shown in Fig. 2A was included to account for the flexibility of the Fab. More precisely, this parameter accounts for variations in the distance between the C-termini of the two Fab $C_H$1 domains to which the DNA was attached due to the following factors: (*i*) the Fab $C_H1$-$C_L$ domains can adopt different conformations with respect to $V_H$-$V_L$ (40) such that the locations of the $C_H1$ C-terminus could shift by up to ≈1 nm; (*ii*) residues C-terminal to $C_H1$ residue 217 were found to be disordered in the 3BNC60 Fab structure (41), thus the position of the $C_H1$ residue to which the DNA was attached (Cys233) is uncertain within ≈1 nm; (*iii*) the ssDNA is covalently linked to the $C_H1$ residue Cys233 using an amine-to-sulfhydryl crosslinker (Sulfo-SMCC), which exerts unknown effects on the length and the degree of flexibility between the ssDNA and Fab.

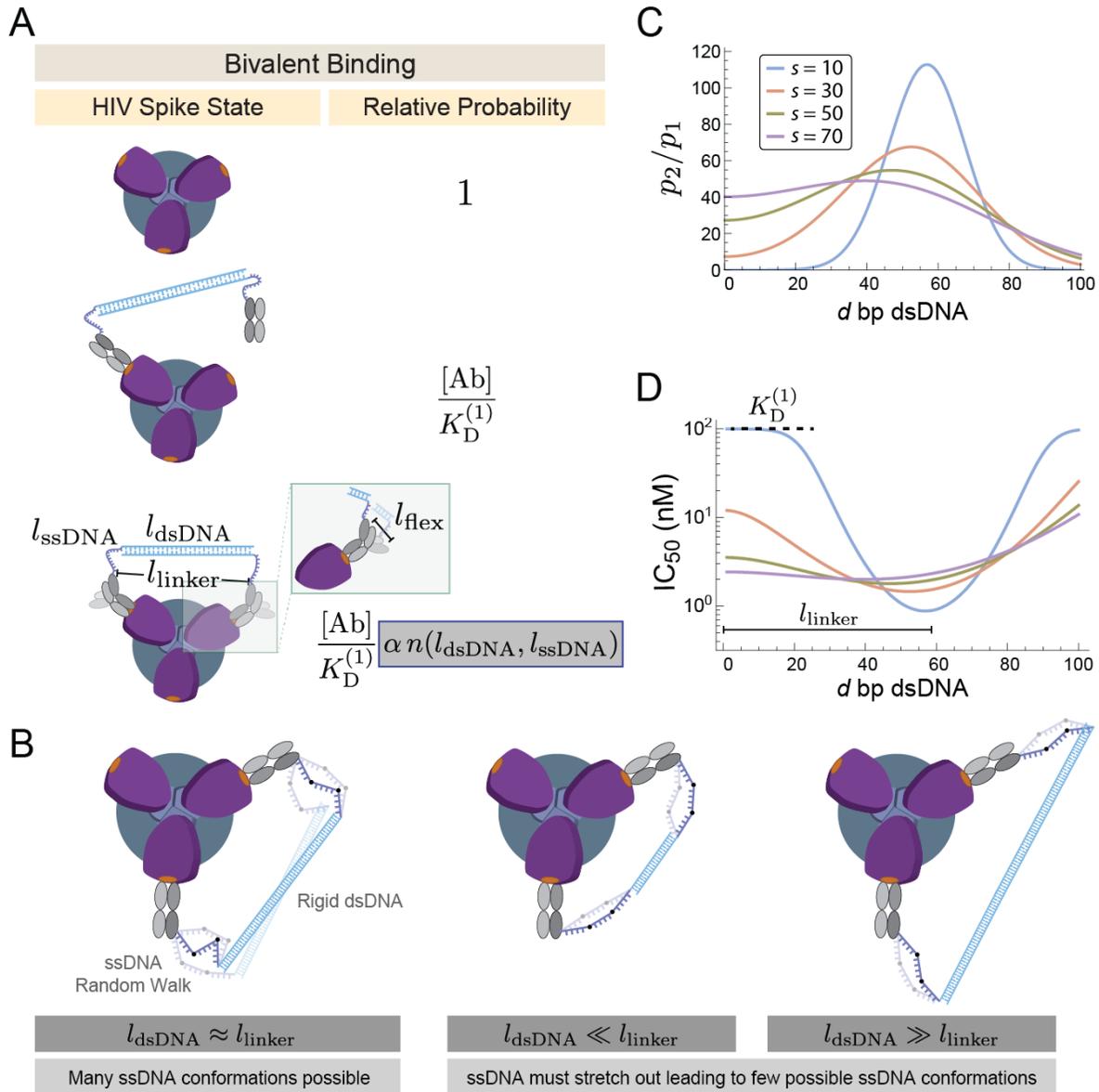

**Fig. 2. Modeling the optimal linker length for a diFab.** (A) The energetic and entropic costs of monovalent binding depend upon the concentration [Ab] and dissociation constant $K_D^{(1)}$ of a single Fab binding. The boost in bivalent binding is dictated by the geometric or avidity factors $\alpha$ (which is the same for all diFabs) and $n$ (which depends upon the length of the dsDNA and ssDNA, the optimal length $l_{\text{linker}}$ of the linker between two bound Fabs, and the flexibility $l_{\text{flex}}$ between the $C_H1$-$C_L$ and $V_H$-$V_L$ domains of a bound Fab). The ability to neutralize is given by the sum of relative probabilities for the monovalent and bivalent states divided by the sum over all states. (B) The boost $\alpha\, n$ of bivalently binding is computed by treating the ssDNA as a random walk and the dsDNA as a rigid rod (Appendix B). An optimal linker matches the length of the dsDNA to the length between two bound Fabs ($l_{\text{dsDNA}} \approx l_{\text{linker}}$), giving rise to many configurations for the bivalently bound state and increasing its likelihood. (C) The relative probability that a diFab with $d$ bp dsDNA and $s$ bases ssDNA is bivalently ($p_2$) versus monovalently ($p_1$) bound and (D) the predicted IC$_{50}$ values for these same constructs. Parameters used were $l_{\text{linker}} = 20$ nm, $l_{\text{flex}} = 1$ nm, $\alpha = 5 \times 10^6$, and $K_D^{(1)} = 100$ nM.

**Relating Neutralization to the Probability that an HIV-1 Spike is Bound.** To model diFab efficacy in terms of the properties of the linker, we first related the ability of a diFab to neutralize an HIV-1 virion to the probability that an Env spike on the surface of HIV-1 will be bound by an antibody. We assume that the spikes are sufficiently far apart to preclude inter-spike crosslinking (Fig. 1A, red boxes) and focus exclusively on intra-spike crosslinking between the three identical sites on the Env homotrimer (Fig. 1A, blue boxes). We further assume that viral infectivity varies linearly with the number of unbound Env, rising from zero (when all spikes are bound by diFabs) to maximum infectivity (when all spikes are unbound) as discussed in Appendix A (42, 43, 44).

Given these assumptions, the ability of a diFab to neutralize HIV-1 is proportional to the probability that at least one of the binding sites on an Env spike will be bound (Appendix A). For example, if each Env protein has a 75% chance to be bound by a diFab, an average of 75% of the spikes on each virion will be bound, and by the linearity assumption, the HIV-1 virions will be 75% neutralized. This enables us to relate the experimentally-determinable % neutralization for diFabs to the theoretically-tractable probability that a single Env spike will be bound either monovalently or bivalently by a diFab. Avidity effects will allow an optimal diFab to bind more tightly to a spike, increasing the binding probability and the neutralization potency.

**The Avidity of a diFab is Dictated by its Linker Composition.** To calculate the probability that any of the Fab binding sites on an HIV-1 spike are occupied, we enumerated three potential states of the spike, which represent a single diFab bound to zero, one, or two binding sites (Fig. 2A). The entropy of the linker was characterized by treating the dsDNA as a 1D rigid rod and the ssDNA as a random walk. The former assumption is valid provided the dsDNA in each linker is less than the 150 bp persistence length of dsDNA (45), a reasonable restriction given that only 60 bp dsDNA are required to span $l_{\text{linker}}$=20 nm. Free ssDNA is flexible with a persistence length $\xi_{\text{ssDNA}} = 1.5$ nm (≈ 2.3 bases) (46, 47) that we analyze using the ideal chain polymer physics model (48).

When a diFab transitions from the unbound state (with probability $p_0$) to a monovalently-bound state (with probability $p_1$), it loses translational and rotational entropy but gains favorable binding energy (49) leading to the relative probability

$$\frac{p_1}{p_0} = \frac{[\text{Ab}]}{K_D^{(1)}}, \qquad [1]$$

where [Ab] is the concentration of the diFab and $K_D^{(1)}$ is the equilibrium dissociation constant of the first diFab arm binding to Env. $K_D^{(1)}$ can be experimentally determined as the IC$_{50}$ (concentration of antibody capable of neutralizing 50% of the virus) of a Fab neutralization profile, with a smaller IC$_{50}$ signifying a more potent antibody. We use the typical value $K_D^{(1)} = $ 100 nM reported for a CD4-binding site bNAb dissociating from a soluble, trimeric HIV-1 Env (50). Importantly, the transition from an unbound to a monovalently-bound diFab is independent of the amount of dsDNA and ssDNA in the diFab's linker.

We now turn to the transition from a monovalently-bound diFab to a bivalently-bound diFab. The ability to simultaneously bind two epitopes depends on the linker composition (the quantity of dsDNA and ssDNA), since the distance $l_{\text{linker}}$ between the C-termini of bivalently-bound Fabs must be spanned by the rigid dsDNA segments as well as the two flanking ssDNA strands (Fig. 2A). More precisely, we enumerate the configurations of the ssDNA random walk and the dsDNA in the bivalently bound state (Fig. 2B), permitting us to compute the entropic cost of bivalent binding (Appendix B). Within this framework, the probability of the bivalently bound state ($p_2$) compared to the monovalently bound state is given by the product of a constant $\alpha$ that is independent of the linker composition and a term $n(l_{\text{dsDNA}}, l_{\text{ssDNA}})$ that depends on the quantity of dsDNA and ssDNA, namely,

$$\frac{p_2}{p_1} = \alpha\, n$$

$$= \frac{\alpha\, l_{\text{flex}}^2}{l_{\text{linker}}\, l_{\text{dsDNA}}} \sqrt{\frac{3}{32\pi^3 l_{\text{ssDNA}} \xi_{\text{ssDNA}}}}\, e^{-\frac{3(l_{\text{linker}}^2 + l_{\text{dsDNA}}^2)}{8 l_{\text{ssDNA}} \xi_{\text{ssDNA}}}} \sinh\left(\frac{3 l_{\text{linker}} l_{\text{dsDNA}}}{4 l_{\text{ssDNA}} \xi_{\text{ssDNA}}}\right) \quad [2]$$

where $l_{\text{dsDNA}} = d\left(0.34\,\frac{\text{nm}}{\text{bp}}\right)$ and $l_{\text{ssDNA}} = s\left(0.64\,\frac{\text{nm}}{\text{base}}\right)$ represent the length of $d$ dsDNA base pairs and $s$ ssDNA bases in the linker, respectively.

Eqs. 1 and 2, together with the normalization condition $p_0 + p_1 + p_2 = 1$, enable us to compute the probability that a diFab will neutralize a virion,

$$p_1 + p_2 = \frac{\frac{[\text{Ab}]}{K_D^{(1)}} + \frac{[\text{Ab}]}{K_D^{(1)}} \alpha\, n}{1 + \frac{[\text{Ab}]}{K_D^{(1)}} + \frac{[\text{Ab}]}{K_D^{(1)}} \alpha\, n}, \quad [3]$$

from which we can write the concentration of 50% inhibition,

$$\text{IC}_{50}^{\text{diFab}} = \frac{K_D^{(1)}}{1 + \alpha\, n}. \quad [4]$$

The value of $\alpha = 5 \times 10^6$ was calibrated from previous measurements where a diFab with $d$=62 bp and $s$=12 bases neutralized HIV-1 approximately 100-fold better than the Fab alone (21). Fig. 2C compares the probability that a diFab will be bivalently bound ($p_2$) rather than monovalently bound ($p_1$) for different linkers. The model shows that bivalent binding is most likely when $l_{\text{linker}} \approx l_{\text{dsDNA}}$, when the rigid dsDNA approximately spans the length between the two bound Fabs. This peak shifts leftwards with the root-mean-squared length of the flexible ssDNA, demonstrating that ssDNA can make up for dsDNA that is slightly too short or too long, provided the flexibility of the ssDNA is taken into account. Fig. 2D shows the corresponding IC$_{50}$s for these constructs. Although adding more flexible ssDNA leads to a broader segment of dsDNA lengths capable of enhanced neutralization, the optimal diFab potency is achieved by including less ssDNA and maximizing the rigidity of the linker.

**The diFab Model Allows Bivalent Binding Only when the dsDNA Length is Approximately Equal to the Length of the Linker it Spans.** To gain a qualitative understanding of our results, we examined Eq. 4 in two limits: near the optimal geometry $l_{\text{dsDNA}} \approx l_{\text{linker}}$ where the ability to bind bivalently is maximum, and far from the optimal limit when the diFab is too short or too long to permit bivalent binding through intra-spike crosslinking.

Near the optimal geometry, HIV-1 neutralization occurs predominantly from the bivalently-bound configuration rather than the monovalent state, $\frac{p_2}{p_1} = \alpha\, n \gg 1$. Hence, the system is well approximated with each spike either being unbound or bivalently-bound, with the dissociation constant $K_D^{(1)}$ for a single Fab boosted by the avidity factor $\alpha\, n$, namely,

$$\text{IC}_{50}^{\text{diFab}} \approx \frac{K_D^{(1)}}{\alpha\, n}. \quad \text{(near optimal geometry)} \tag{5}$$

If the potency of diFab 1 is $\text{IC}_{50}^{(1)}$, and the potency of diFab 2 with a different linker is $\text{IC}_{50}^{(2)}$, the latter diFab's potency will be shifted relative to the former by the ratio of *n* factors, namely,

$$\frac{\text{IC}_{50}^{(2)}}{\text{IC}_{50}^{(1)}} = \frac{n^{(1)}}{n^{(2)}}. \tag{6}$$

In other words, the relative potency of both diFabs is determined solely by the entropy, rather than the energy, of the linker when bivalently bound.

In the opposite regime where the diFab linker is too small ($l_{\text{dsDNA}} + l_{\text{ssDNA}} \lesssim l_{\text{linker}}$) or too large ($l_{\text{dsDNA}} - l_{\text{ssDNA}} \gtrsim l_{\text{linker}}$), the diFab loses the ability to bind bivalently and the IC$_{50}$ attains a constant value

$$\text{IC}_{50} \approx K_D^{(1)} \quad \text{(far from optimal geometry)} \tag{7}$$

shown as a black dashed line in Fig. 2D.

**The Avidity of a triFab is Capable of Binding Three Env Sites is Further Enhanced over that of the diFab.** While the diFab constructs were inspired by two-armed IgG antibodies, a triFab construct that could simultaneously bind to three Env epitopes (Fig. 1B) should exhibit even greater avidity and hence neutralize HIV-1 more potently. For simplicity, we assume that both dsDNA segments have the same length, as do all ssDNA segments. As derived in Appendix C, the IC$_{50}$ of such a construct is given by

$$\text{IC}_{50}^{\text{triFab}} = \frac{K_D^{(1)}}{1 + \alpha\, n + \alpha^2 n^2}. \tag{8}$$

Fig. 3A and B compare the neutralization potencies across the design space of diFabs and triFabs, demonstrating that joining three Fabs can result in IC$_{50}$s far smaller than what is possible for even the theoretically optimal diFab design.

The neutralization potency of a triFab near its optimal geometry ($\alpha\, n \gg 1$) is dictated purely by the trivalently bound state,

$$\text{IC}_{50}^{\text{triFab}} \approx \frac{K_D^{(1)}}{\alpha^2 n^2}. \quad \text{(near optimal geometry)} \tag{9}$$

Note that the boost in avidity in going from a Fab to a diFab is equivalent to the boost between a diFab and triFab. Since diFabs have been shown to achieve a 100-1000 fold decrease in IC$_{50}$ over a Fab, a triFab should be able to achieve a 10$^4$-10$^6$ fold decrease in IC$_{50}$ relative to the Fab, providing a powerful framework with which to achieve very high neutralization. To further enhance potency, more than three Fabs could be joined together into a multiFab. By joining multiple types of Fabs that bind to different epitopes (e.g., the broadly neutralizing CD4-binding 3BNC117 with a V1V2-binding PG16), multiFabs may enhance the potency of neutralization as well as help combat HIV-1 heterogeneity as has been seen in combination influenza antibodies (51) and HIV-1 mosaic vaccines (52).

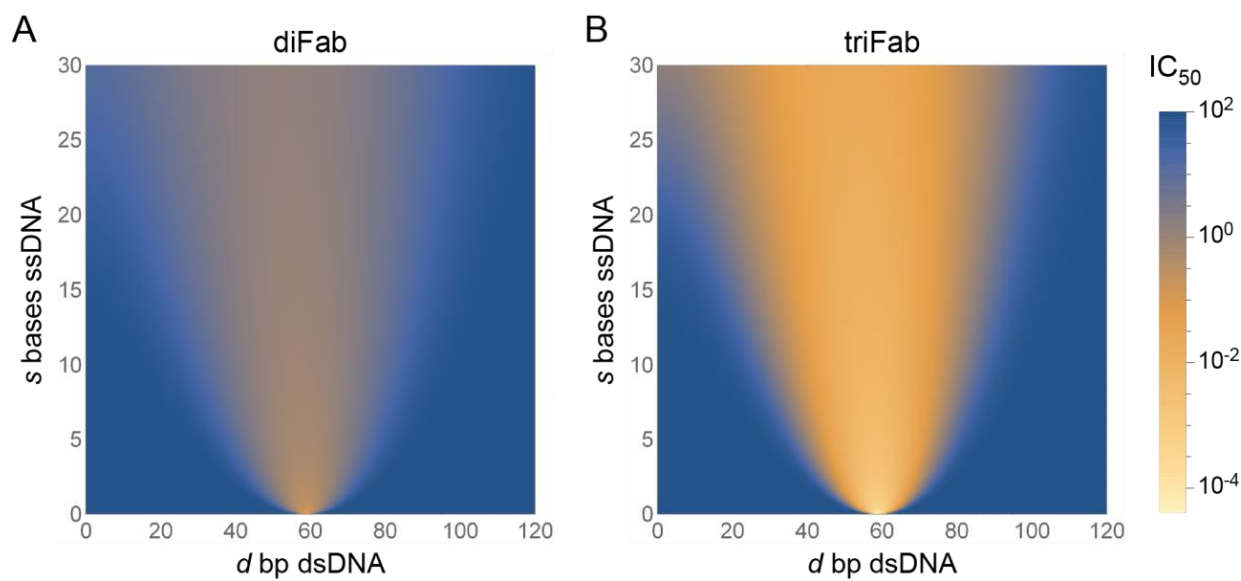

**Fig. 3. Neutralization potency of different diFab and triFab designs.** Predicted IC$_{50}$s for (A) diFab or (B) triFab constructs with linkers composed of $d$ bp dsDNA and $s$ bases ssDNA segments. In both cases, optimal neutralization occurs when the linker matches the length between Fab binding sites ($d = \left(3\frac{\text{bp}}{\text{nm}}\right) l_{\text{linker}} = 60$ nm) and has as little ssDNA as possible. The triFab can achieve significantly enhanced potency over the diFab because of its additional avidity.

**An Optimal Linker is Maximally Rigid and Perfectly Spans the Distance between Env Epitopes.** Counter to what might be intuitively expected, adding flexibility to a linker by increasing the number of ssDNA bases need not improve the diFab's neutralization potential. This effect arises because flexible ssDNA has a large number of degrees of freedom that are constrained when the diFab bivalently binds, leading to a larger entropic penalty (or smaller boost in avidity). Indeed, the full diFab and triFab design space shown in Fig. 3 demonstrates that an optimal construct is a perfectly rigid linker composed of only dsDNA whose length matches the distance between HIV-1 epitopes ($l_{\text{dsDNA}} \approx l_{\text{linker}}$). However, we caution that such diFabs may not operate optimally for both experimental and theoretical reasons including: (1) the range of dsDNA lengths that permit bivalent binding narrows as the amount of ssDNA decrease, and in the extreme limit of a linker with no ssDNA, being slightly too short or too long by as little as a few base pairs may preclude bivalent binding; (2) charge interactions between the dsDNA and either the Fab or Env may disrupt diFab functionality; (3) a lack of sufficient flexibility at the Fab-dsDNA junction may preclude bivalent binding; and (4) in the limit of a rigid linker (ssDNA≲10 bases) the ideal chain model breaks down and the bending of ssDNA must be accounted for (e.g., via the worm-like chain model) which may result in higher IC$_{50}$s than predicted by our model.

Because of these considerations, it is worthwhile to use a diFab whose dsDNA length $l_{\text{dsDNA}}$ matches the distance between Fab epitopes ($l_{\text{linker}}$) found in crystal structures, and to flank this dsDNA with short segments of ssDNA (≈10 bases) to buffer against uncertainties in the measurements. As seen in Fig. 3, such constructs lie within a wide basin with strong neutralization potential. Of note, this strategy far surpasses a purely flexible linker ($l_{\text{ssDNA}} \approx l_{\text{linker}}$, $l_{\text{dsDNA}} = 0$) which cannot bivalently bind because of the large entropic cost. Other methods of introducing flexibility, such as introducing ssDNA breaks in the dsDNA linker, are also predicted to increase the entropic cost of bivalent binding and hence increase the IC$_{50}$ (see Appendix C).

**Discussion**

The low density of Env spikes on HIV-1 potentially enables the virus to mitigate the host antibody response by hindering IgGs from using both antigen-binding Fabs to bind bivalently, thereby expanding the range of HIV-1 mutations permitting antibody evasion (21, 22). Indeed, a mutant simian immunodeficiency virus (SIV) with a higher number of Env spikes reverted to its normal spike count of ≈14 when propagated in non-human primate hosts (53). This suggests that while HIV-1 may be more infectious with more Env trimers (54), the immune system applies selective pressure that keeps the Env spike count per virion low, presumably to prevent anti-HIV-1 IgGs from utilizing avidity effects to counter the lower intrinsic Fab-Env affinities that result from rapid mutation of Env. Antibody isotypes such as dimeric IgA or pentameric IgM have increased valencies (four and ten Fabs, respectively) compared to the two Fabs of an IgG, thus allowing for increased avidity effects during antibody binding to a pathogen. However, most of the neutralizing activity in the sera of HIV-1–positive individuals is attributed to IgGs (55, 56), and converting an anti-HIV-1 IgG antibody to an IgA or IgM has minimal effects on potency in standard neutralization assays (57, 58), possibly because there are so few Env spaced sufficiently far apart that these other antibody classes are also forced to bind monovalently.

Bivalent binding to single Env trimer (intra-spike crosslinking) is another way to utilize avidity effects to counteract the low spike density of HIV-1. Although the architecture of Env trimers prohibits this mode of binding for conventional, host-derived IgGs (29, 30), we analyzed how synthetic diFabs (Fabs from a neutralizing anti-HIV-1 IgG joined by a linker containing rigid dsDNA flanked by flexible ssDNA shown in Fig. 1B) could be designed to achieve optimal intra-spike crosslinking.

HIV-1 Env trimers adopt multiple conformations on virions (24, 59) and in the soluble native-like forms used for structural studies (60). For example, binding of the host CD4 receptor induces outward displacements of the three Env gp120 subunits, resulting in an open conformation in which the coreceptor binding sites on the trimer apex V3 loops are exposed (38, 39, 61, 62) and that rearranges further upon coreceptor binding and subsequent membrane fusion. We measured the distances between adjacent 3BNC117 epitopes in a new cryo-EM structure (63) and estimated the average position of the C-terminal C$_H$1 domain residue to which the DNA of a 3BNC117 diFab would be covalently attached (40). Based on these measurements, we assumed that a diFab neutralized HIV-1 when the linker spans a length $l_{\text{linker}}$=20 nm. We further assumed that the Fab C$_H$1-C$_L$ domains can stretch relative to the V$_H$-V$_L$ domains by $l_{\text{flex}}$ =1 nm (see Fig. 2A). Lastly, we considered a Fab whose ability to dissociation (as given by the midpoint of a neutralization assay) equals to $K_D^{(1)}$ =100 nM. Each of these values can be readily adapted to other HIV-1 strains and Fabs.

With these parameters in hand, we developed a statistical mechanics-based model to predict the neutralization of a diFab whose linker is composed of $d$ base pairs dsDNA and $s$ bases ssDNA (Fig. 1B), enabling us to tune both the length and rigidity of the linker. By assuming that (*i*) each homotrimeric spike is unable to help infect a host cell when any one of its three epitopes are bound by Fab and (*ii*) that the infectivity of a virion varies linearly with the number of unbound Env, we showed that the neutralization of a virion is proportional to the probability that any single Env protein is bound by a Fab (Appendix A). This framework enabled us to translate the linker-dependent entropy and energy of binding to an HIV-1 Env trimer into the predicted neutralization potency for each diFab.

It is worthwhile to point out several factors that the model neglects. First, the model does not consider potential, but presumably rare, diFab binding between adjacent Env trimers. Second, our model assumed that the % neutralization of HIV-1 decreases linearly with the number of unbound Env trimers (Appendix A). While such a linear relationship was observed when less than half of HIV-1 spikes were bound (54), it may break down if, for example, at least 2-3 unbound Env trimers are needed to infect a cell. We also assumed that each virion had exactly 14 spikes, neglecting the relatively small observed spike number variations (23-26). However, relaxing these assumptions yielded nearly identical results (Appendix A), suggesting that our results are robust and should apply to more general descriptions of HIV-1 neutralization.

We determined that an optimal linker will maximize its rigidity, trading flexible ssDNA for rigid dsDNA. The larger flexibility of ssDNA implies that there will be a higher entropic penalty for bivalently binding, thereby resulting in a worse (larger) $IC_{50}$. This general statement applies to all forms of increased flexibility including: (1) trading dsDNA for an equivalent length of ssDNA; (2) adding ssDNA without decreasing the length of dsDNA; or (3) introducing ssDNA gaps in the dsDNA (Appendix C).

In additional to tuning the length of dsDNA and ssDNA in the linker, an additional biologically-inspired approach to further increase avidity is to construct multiFabs that target more than two epitopes. As has been seen in other biological systems (64-66), higher valencies can elicit tighter binding. Hence, a triFab that allows three Fabs to simultaneously bind to an HIV-1 Env trimer is predicted to have a lower $IC_{50}$ than an optimal diFab; indeed, our model predicts that the boost from avidity of an optimal diFab over a Fab is equivalent to the boost of an optimal triFab over an optimal diFab, and hence we expect that triFabs should be able to achieve $IC_{50}$s $10^4$-$10^6$ fold smaller than Fabs. MultiFabs that link together Fabs targeting different epitopes (e.g., the broadly neutralizing CD4-binding site antibody 3BNC117 and the V2-binding PG16 antibody) (52) could better combat the heterogeneity of HIV-1 strains, providing guidance for constructing optimal anti-HIV-1 therapeutics that remain potent against HIV-1 in the face of the Env mutations arising during HIV-1 replication.

## Supporting Information

The supporting information includes appendices and a Mathematica notebook that reproduces the figures in this manuscript.

## Author Contributions

T.E., A.P.W., R.P. and P.J.B. conceived the project; T.E., S.Y., and R.P. developed the model and performed analyses; T.E., R.P., and P.J.B. wrote the paper with input from other authors.

## Acknowledgements
We thank Anthony Bartolotta, Justin Bois, Jim Eisenstein, Vahe Galstyan, Peng He, Willem Kegel, David Hsieh, Giacomo Koszegi, Pankaj Mehta, Jiseon Min, Olexei Motrunich, Noah Olsman, Vahe Singh, and Richard Zhu for useful discussions, Christopher Barnes for measuring modeled 3BNC60-Env complexes, and Marta Murphy for help preparing figures. This research was supported by NIH NIAID grants 1R01AI129784 and HIVRAD P01 AI100148 (P.J.B.), the Bill and Melinda Gates Foundation Collaboration for AIDS Vaccine Discovery Grant 1040753 (P.J.B.), La Fondation Pierre-Gilles de Gennes (R.P.), the Rosen Center at Caltech (R.P.), R01

GM085286, and 1R35 GM118043-01 (MIRA) (R.P.), and a Caltech-COH Biomedical Research Initiative (P.J.B.). We thank the Burroughs-Wellcome Fund for their support through the Career Award at the Scientific Interface (S.Y.) as well as for the Physiology Course at the Marine Biological Laboratory where part of this work was done.

# Supporting Information

## A. % Neutralization and the Probability that HIV-1 Envelope is Bound

In this section, we determine the relationship between the experimentally measurable % neutralization and the probability $p_{\text{bound}}$ (denoted by $p_1 + p_2$ in the main text) that a trimeric Env spike will have a Fab bound to any of its three identical epitopes.

We begin by defining the typical % neutralization assay. We then analyze a linear model of HIV-1 infectivity discussed in the main text where the ability of the virus to infect a target cell is proportional to the number of Env spikes not bound by an antibody. We also discuss alternative models of infectivity in which some minimal number of active spikes is required for a virion to infect a cell, finding that this model yields nearly identical predictions to the linear model.

Lastly, we investigate the importance of the experimentally-measured distribution of Env spikes on HIV. We start by assuming that each virion has the same number of spikes (given by the mean of the measured distribution) and then relax this assumption to characterize how infectivity changes when the number of Env spikes per virion is drawn from the full distribution.

**Defining % Neutralization**

In previous work, in vitro neutralization assays were carried out in 96 well plates, each well containing 250 TCID$_{50}$ and 25,000 cells that emit bioluminescence upon infection by the pseudovirus as described (67). Antibodies and other potential inhibitors of neutralization are in vast excess over pseudovirus and cells in these assays (e.g., 1 nM antibody corresponds to $10^{11}$ molecules/well). Upon infection, the cells emit light via a luciferase reporter as shown in Fig. S1A. We define the percent of pseudovirus neutralized as the fold-change in bioluminescence in the presence and absence of an inhibitor, namely,

$$\% \text{ neutralization} = 100 \frac{(\text{viral control} - \text{cell control}) - (\text{bioluminescence} - \text{cell control})}{\text{viral control} - \text{cell control}}, \quad [S1]$$

where *bioluminescence* is a measure of the light emitted in a well containing the pseudoviruses, cells, and antibodies; *viral control* is an assay using only cells and viruses (no antibodies) so that the cells emit maximal bioluminescence; and *cell control* is an assay using only cells (no viruses or antibodies) (Fig. S1A).

The relative infectivity of a virion is defined as $100 - (\% \text{ neutralization})$, which equals 100% when no antibodies are present and 0% at saturating antibody concentrations where the binding sites on each HIV-1 spike are occupied.

From these assays, we define the equilibrium dissociation constant $K_D^{(1)}$ characterizing the binding of the first diFab arm to HIV-1 Env (which is equivalent to the dissociation constant between a diFab with only one functional arm) as the midpoint of the neutralization curve. In this work, we chose the typical value $K_D^{(1)} = 100$ nM from previous experiments.

**A Linear Model of HIV-1 Infectivity**

We now consider the linear model for HIV-1 infectivity used in the main text, which is predicated on the following assumptions: (*i*) each virus has the same number $N = 14$ of Env trimers (or spikes), taken to be the mean of the experimentally measured distribution, (*ii*) each spike is active (able to help HIV-1 infect a target cell) if and only if none of its three subunits are bound by an antibody, and (*iii*) the relative infectivity of a virion is linearly proportional to its number of active spikes. Taken together, these assumptions imply that a virus with 7 active Env trimers shown on the left in Fig. S1B will be half as infective as a completely unbound virus with 14 active trimers shown on the right.

The first assumption simplifies our analysis; below we relax this assumption and show that it minimally alters the % neutralization curves. The second assumption, that a spike is inactivated if at least one of its Fab sites is bound, is supported by experimental and computational studies (23). The third assumption has been observed when less than half the HIV-1 spikes are bound (27), but relative infectivity decreases faster than a linear model when more than half of the spikes are bound. In the following section, we relax this assumption and show that it minimally alters the % neutralization curves by making them slightly sharper. Taken together, these calculations demonstrate that our results are robust to the details of HIV-1 neutralization.

With these assumptions, a virion with $n$ active spikes out of $N = 14$ spikes total will have $\frac{n}{N}$ the relative infectivity of a completely unbound virus, and hence the % neutralization is given by the expectation $100 \langle 1 - \frac{n}{N} \rangle$, which equals 0% in the absence of antibodies when all spikes are active ($n = N$) and equals 100% at saturating antibody concentrations when all spikes are inactive ($n = 0$). Given the probability $p_{\text{bound}}$ that any spike will be bound, the probability of having $n$ active (unbound) spikes equals $p(n) = \binom{N}{n}(1 - p_{\text{bound}})^n p_{\text{bound}}^{N-n}$ and hence % neutralization is given by the weighted average of having $n$ unbound spikes,

$$\begin{aligned} \% \text{ neutralization} &= 100 \sum_{n=0}^{N} p(n)\left(1 - \frac{n}{N}\right) \\ &= 100 \sum_{n=0}^{N} \binom{N}{n}(1 - p_{\text{bound}})^n p_{\text{bound}}^{N-n}\left(1 - \frac{n}{N}\right) \\ &= 100 - \frac{100}{N} \sum_{n=0}^{N} \binom{N}{n}(1 - p_{\text{bound}})^n p_{\text{bound}}^{N-n} n \\ &= 100 - \frac{100}{N} N(1 - p_{\text{bound}}) \\ &= 100 \, p_{\text{bound}}, \end{aligned} \quad [\text{S2}]$$

where in the fourth equality we used the average value of the binomial distribution. Therefore, the statistical mechanical model we develop to characterize the probability that a diFab is bound to Env also describes the ability of that diFab to neutralize a virion.

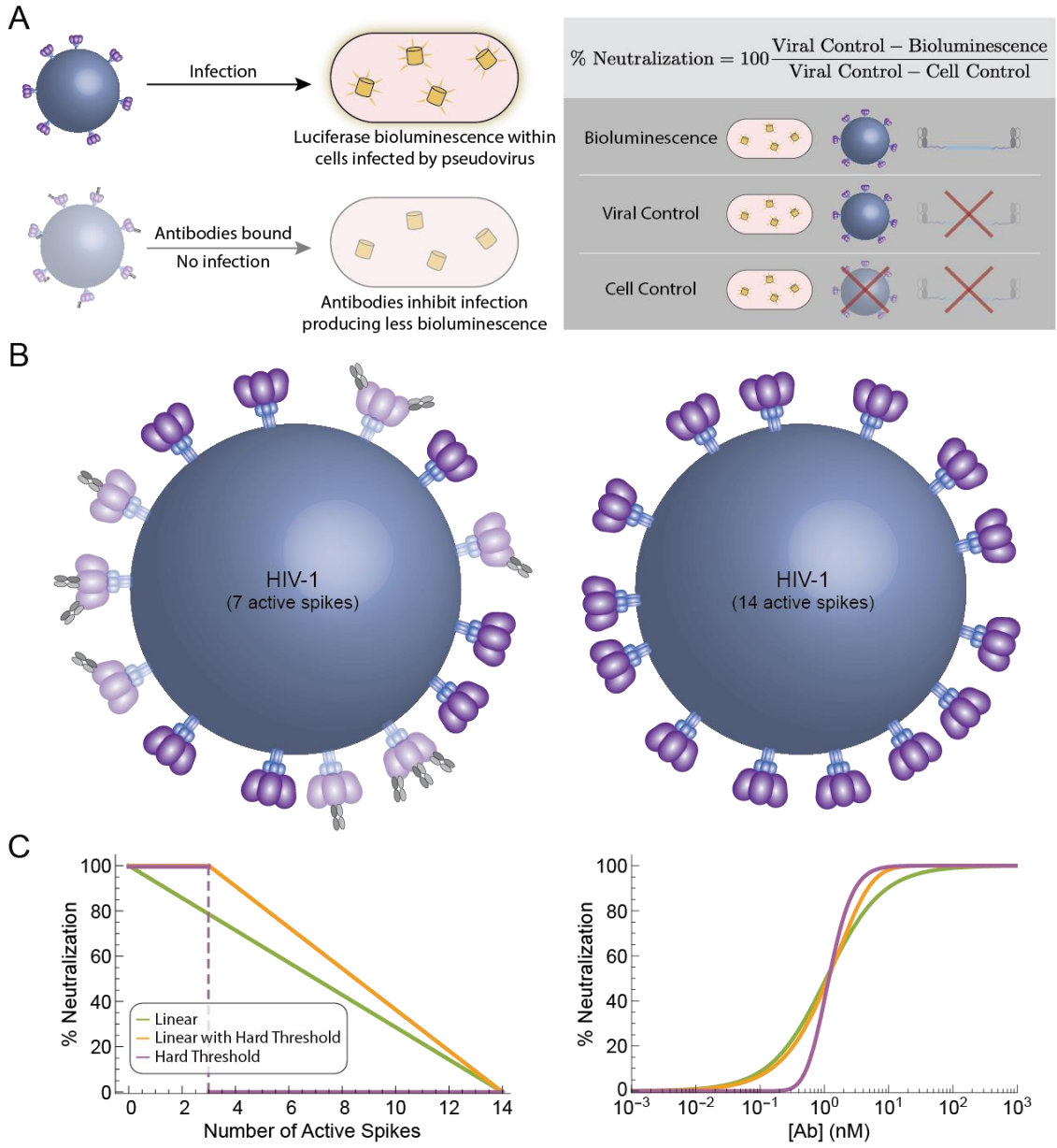

**Figure S1. A Linear model of HIV-1 neutralization.** (A) In previous work, neutralization of HIV-1 pseudovirus was assayed by evaluating reporter cells that emit light upon HIV-1 infection via luciferase (1). Adding an antibody results in decreased bioluminescence. (B) We model each HIV-1 virion as having 14 spikes that are inactivated (represented as partially transparent spikes) when a Fab is bound to any of its three binding sites on an Env trimer. In the main text, we assume a linear model in which the infectivity of a virion is proportional to its number of active spikes; for example, the virus on the right will be twice as infective as the virus on the left. (C) Comparison of different models for % neutralization (or the relative infectivity defined as $100 - \%\ \text{neutralization}$) as a function of the number of active (unbound) HIV-1 spikes (left) or antibody concentration (right). The geometric factor $\alpha$ in Eq. 4 quantifying the effects of diFab avidity was adjusted for each model ($5 \times 10^6$ for the linear model; $3 \times 10^6$ for the linear model with a hard threshold; $20 \times 10^6$ for the hard threshold model) to match their IC$_{50}$s. The values of the remaining parameters were the same as in Fig. 2.

**Imposing a Hard Threshold for HIV-1 Infectivity**
We now relax the third assumption stated above that relative infectivity is proportional to the number of active (i.e., unbound) HIV-1 spikes (Fig. S1C, *Linear*). Instead, we posit that some minimum number of spikes must be active for a virion to be able to infect a target cell. This minimum number has been predicted to be between 1-3 active spikes (42). Hence, we investigate two additional models where at least 3 of the HIV-1 spikes must be active for a virion to infect a target cell. In the first model (Fig. S1C, *Linear with a Hard Threshold*), the relative infectivity increases linearly (and hence the % neutralization decreases linearly) with the number of active spikes >3, while in the second model (Fig. S1C, *Hard Threshold*) we impose a pure threshold so that a virus is maximally infective provided at least 3 spikes are active.

For each model, we can alter the avidity factor $\alpha$ to match the midpoints of each curve ($\alpha = 5 \times 10^5$, $3 \times 10^5$, or $20 \times 10^5$ for the linear model, the linear model with a hard threshold, or the hard threshold model, respectively), since this parameter would ordinarily be inferred from a neuralization assay. As shown in Fig. S1C, the linear model with a hard threshold is nearly identical to the linear model without this threshold, except that its % neutralization rises to 100% at lower antibody concentrations because it only needs to neutralize $N-2$ spikes to disable each virion. The hard-threshold model is sharper than the linear model, with the transition between no neutralization and full neutralization occurring when there are enough antibodies to bind $N-2$ spikes. Having noted these slight discrepancies, using any of these models would minimally affect our results, and hence we chose to proceed using the simplest linear response.

**% Neutralization is Unchanged if the Number of Env Spikes Varies between Virions**
In this section, we relax the first assumption stated above and consider the number of spikes $N$ on each virion to be drawn from a distribution ranging from 7-30 spikes per virion with an average of 14 (22-26). We assume that the relative infectivity of a virus increases with each additional spike (with a maximum value attained by a virus with $N_{\max} = 30$ active spikes). The calculation for % neutralization follows analogously to Eq. S2, except that the % neutralization of a virus with $n$ active spikes is proportional to $1 - \frac{n}{N_{\max}}$ and that % neutralization must be averaged over all possible values of $N$ drawn from its distribution, namely,

$$\% \text{ neutralization} \propto \langle \sum_{n=0}^{N} \binom{N}{n} (1 - p_{\text{bound}})^n p_{\text{bound}}^{N-n} \left(1 - \frac{n}{N_{\max}}\right) \rangle$$
$$= 1 - \frac{\langle N \rangle}{N_{\max}}(1 - p_{\text{bound}}). \qquad [S3]$$

In experimental measurements, % neutralization is always stretched to run from 0% to 100%, and if we apply this same stretching to Eq. S3 we recover the result that $\% \text{ neutralization} = 100 p_{\text{bound}}$ as in Eq. S2.

This result is noteworthy in that it is independent of any details of the distribution (even the mean of the distribution does not affect neutralization when the function is stretched to run from 0% to 100%, as is customary practice in experimental measurements). However, this calculation rests upon the underlying assumption that relative infectivity increases linearly with the number of active spikes with no upper bound. If the relative infectivity saturates past a certain point (e.g., following a sigmoidal dependence), the result would no longer hold. Yet given that HIV-1 has so few spikes, it may turn out that each spike increases the relative infectivity of the virus by the same comparable amount.

## B. Number of Microstates in the Model Including ssDNA

In this section, we enumerate the number of microstates associated with the three conformations of an HIV-1 Env spike – the unbound ($\Omega_0$), singly bound ($\Omega_1$), and bivalently bound states ($\Omega_2$) – shown in Fig. S2. Our goal is to quantify the small fraction of linker configurations that enable a diFab to span the distance $l_{\text{linker}}$ and bivalently bind.

The relative probability of each Env state is proportional to its Boltzmann weight, $e^{-\beta(E_j - TS_j)} = \Omega_j e^{-\beta E_j}$, where $\beta = \frac{1}{k_B T}$, $E_j$ is the energy of state $j$, $T$ is the temperature (37˚C), $k_B$ = Boltzmann's constant, $S = k_B \log \Omega_j$ is the entropy, and $\Omega_j$ is the number of microstates of state $j$. Thus, the relative probabilities of the bivalent and monovalent states satisfy

$$\frac{p_2}{p_1} = \frac{\Omega_2}{\Omega_1} e^{-\beta(E_2 - E_1)} \quad \text{[S4]}$$

where $p_j$, $E_j$, and $\Omega_j$ represent the probability, energy, number of microstates when $j$ Fabs are bound to an Env, respectively.

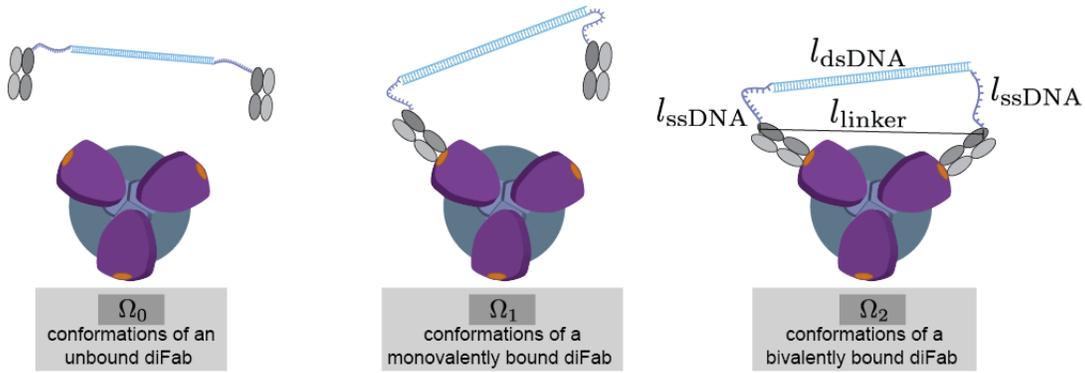

**Figure S2. Conformational states of a diFab.** We denote the number of microstates $\Omega_j$ of a diFab bound with $j$ Fabs. To bivalently bind, the linker region of a diFab (a length $l_{\text{dsDNA}}$ of dsDNA flanked by two lengths $l_{\text{ssDNA}}$ of ssDNA) must span a distance $l_{\text{linker}}$ that is dictated by the Env's structure. The dsDNA is a rigid rod while the ssDNA is modeled as an ideal chain.

The number of configurations $\Omega_1$ for a monovalently-bound diFab is proportional to the number of degrees of freedom available when one Fab is bound to its epitope and the second Fab is unbound but tethered to the first by DNA. More precisely, the different conformations arise from the rotation of the rigid dsDNA together with the small ssDNA motions. Since the associated entropy of this motion increases slowly with size, $\Omega_1$ is effectively constant across all diFabs regardless of their linker composition. Similarly, since we assume that all diFabs are made up of the same Fab (e.g., all made using 3BNC117), the binding energy $E_2 - E_1$ gained when the second Fab in a diFab binds to the HIV-1 spike is the same across all constructs. Thus, the only parameter in Eq. S4 that varies across diFabs with different linkers is the number of states $\Omega_2$ of a bivalently bound linker, which we now proceed to compute.

### The Ideal Chain Model for ssDNA and dsDNA
We model the dsDNA as a 1D rigid rod and the ssDNA as a random walk with a step size given by its Kuhn length $b_{\text{ssDNA}} = 2\xi_{\text{ssDNA}} = 3$ nm = 4.7 bases (46, 47). As in the main text, $l_{\text{dsDNA}}$ denotes the total length of dsDNA while $l_{\text{ssDNA}}$ denote the length of each of the two ssDNA segments.

Intuitively, because random walks tend to wander around their starting point, the optimal diFab will match its dsDNA length to the distance between the C-termini of two bound Fabs ($l_{\text{dsDNA}} \approx l_{\text{linker}}$) to lose as little entropy as possibly when transitioning from a monovalently-bound to a bivalently-bound state. As shown in Fig. 2B of the main text, a diFab that is too short ($l_{\text{dsDNA}} \ll l_{\text{linker}}$) or too long ($l_{\text{dsDNA}} \gg l_{\text{linker}}$) must stretch its ssDNA outwards or inwards to bivalently bind, thereby severely limiting the number of possible configurations in the doubly bound state. In the extreme limits where $l_{\text{linker}} > l_{\text{dsDNA}} + 2l_{\text{ssDNA}} + l_{\text{flex}}$ or $l_{\text{linker}} < l_{\text{dsDNA}} - 2l_{\text{ssDNA}} - l_{\text{flex}}$, bivalent binding is impossible (recall that $l_{\text{flex}}$ is the length by which the Fab can stretch as shown in Fig. 2A). We will now make these statements precise by computing the probability that a linker configuration will permit a diFab to be bivalently bound.

**Computing the Probability of Bivalent Binding**

We first turn our attention to the number of microstates of a bivalently bound linker ignoring the flexibility of the Fab shown in Fig. 3A (i.e., in the $l_{\text{flex}} = 0$ limit). Our goal will be to compute the probability $p(l_{\text{dsDNA}}, l_{\text{ssDNA}}, l_{\text{linker}})$ that the two ssDNA random walks and the dsDNA segment will span the appropriate distance $l_{\text{linker}}$ necessary for the two Fabs to bivalently bind and then use this probability to count the number of microstates available for bivalent binding. In the last section of this Appendix, we consider the case where $l_{\text{flex}} \neq 0$.

As shown in Fig. S3A, each ssDNA random walk is composed of $\hat{n} = \frac{l_{\text{ssDNA}}}{b_{\text{ssDNA}}}$ segments (where the notation distinguishes this variable from $n$ in Eq. 2) with Kuhn length $b_{\text{ssDNA}} = 2\xi_{\text{ssDNA}}$. The ssDNA and dsDNA in the linker must together span a fixed $\vec{l}_{\text{linker}}$, where the direction and magnitude of this vector is determined by the geometry of the Env spike's epitopes. We now compute the probability that two ssDNA random walks sandwiched between a dsDNA rigid rod of size $\vec{l}_{\text{dsDNA}}$ spans $\vec{l}_{\text{linker}}$. We proceed by considering four increasingly complex cases.

Case 1: $\vec{l}_{\text{dsDNA}} = \vec{0}, \vec{l}_{\text{linker}} = \vec{0}$

We begin by analyzing the special case of a diFab with an infinitesimally small dsDNA segment ($\vec{l}_{\text{dsDNA}} = \vec{0}$) binding to two epitopes that essentially lie on top of one another ($\vec{l}_{\text{linker}} = \vec{0}$). In other words, the constraint $\vec{l}_{\text{dsDNA}} = \vec{0}$ implies that both ssDNA random walks start off at the same location whereas $\vec{l}_{\text{linker}} = \vec{0}$ specifies that both random walks must end at the same location. This setup is shown in Fig. S3B where the two random walks begin in the green square and end up within a small distance of each other represented by the gray cube.

Rather than analyzing the first random walk (with steps $\vec{s}_1(1), \vec{s}_1(2)$, and $\vec{s}_1(3)$) and the second random walk (with steps $\vec{s}_2(1), \vec{s}_2(2)$, and $\vec{s}_2(3)$) individually, we construct an effective random walk that traverses along one of the original walks and back along the other ($\vec{s}_{\text{eff}}(j) = \vec{s}_1(j)$ for $1 \leq j \leq 3$ and $\vec{s}_{\text{eff}}(j) = -\vec{s}_2(j-3)$ for $4 \leq j \leq 6$). This mapping is bijective, which means that every instance of the original random walks will correspond to a unique effective random walk and vice versa. Hence, the two original random walks will end at the same point if and only if the effective random walk ends near the origin. Therefore, $p(0, l_{\text{ssDNA}}, 0)$ equals the probability that this effective random walk returns to the origin.

To make this argument precise, consider a 3D random walk starting at the origin and taking $\hat{n}$ steps of length $b_{\text{ssDNA}}$. The probability that a random walk will end inside an infinitesimal volume $dV$ centered at $\vec{r}$ is given by (69)

$$P(\vec{r})dV = dV\left(\frac{3}{2\pi\hat{n}b_{\text{ssDNA}}^2}\right)^{\frac{3}{2}} e^{-\frac{3r^2}{2\hat{n}b_{\text{ssDNA}}^2}}. \qquad [S5]$$

Since each segment of the random walk can point in any direction, this probability only depends on the magnitude of $\vec{r}$ and decreases exponentially with its distance from the origin. Note that $P(\vec{r})$ is a probability density that upon multiplication by an infinitesimal volume $dV$ denotes the probability of a random walk ending between $\vec{r}$ and $\vec{r} + d\vec{r}$.

The probability that the effective random walk, formed by combining the two ssDNA random walks, will end up inside an infinitesimal volume $dV$ around the origin is given by Eq. S5 with $\hat{n} \to 2\hat{n}$ and $\vec{r} \to \vec{0}$. Therefore, the fraction of bivalent binding configurations relative to monovalent binding configurations for two ssDNA random walks that start and end at the same location ($\vec{l}_{\text{dsDNA}} = \vec{0}, \vec{l}_{\text{linker}} = \vec{0}$) is given by

$$p(0, l_{\text{ssDNA}}, 0) = P(\vec{0})dV = dV\left(\frac{3}{4\pi\hat{n}b_{\text{ssDNA}}^2}\right)^{\frac{3}{2}}. \qquad [S6]$$

Thus far, our calculation has been in terms of the probability $P(\vec{0})dV$. To convert this result into the number of microstates (and thereby compute the entropy), define $\bar{\Omega}$ to be the number of microstates of each independent ssDNA segment (the $2\hat{n}$ ssDNA segments are all assumed to rotate freely in the ideal chain model), the total number of microstates for the bivalent binding configurations is given by $\bar{\Omega}^{2n}p(0, l_{\text{ssDNA}}, 0)$. We note that this $\bar{\Omega}$ will ultimately be subsumed into the $\alpha$ factor in Eq. 2.

Case 2: $\vec{l}_{\text{dsDNA}} = \vec{0}, \vec{l}_{\text{linker}} \neq \vec{0}$
We next consider the case where two ssDNA random chains with no interspersed dsDNA ($\vec{l}_{\text{dsDNA}} = 0$) must end up at a displacement $\vec{l}_{\text{linker}}$. As shown in Fig. S3C, the two ssDNA random walks are equivalent to the probability that a single random walk with $2\hat{n}$ steps will finish at $\vec{l}_{\text{linker}}$. Therefore, the fraction of random walk configurations that allow bivalent binding when $\vec{l}_{\text{dsDNA}} = \vec{0}$ and $\vec{l}_{\text{linker}} \neq \vec{0}$ is given by Eq. S5 with $\hat{n} \to 2\hat{n}$ and $\vec{r} \to \vec{l}_{\text{linker}}$, namely,

$$p(0, l_{\text{ssDNA}}, l_{\text{linker}}) = P(\vec{l}_{\text{linker}})dV = dV\left(\frac{3}{4\pi\hat{n}b_{\text{ssDNA}}^2}\right)^{\frac{3}{2}} e^{-\frac{3l_{\text{linker}}^2}{4\hat{n}b_{\text{ssDNA}}^2}}. \qquad [S7]$$

As in Case 1, the total number of bivalent microstates is then given by $\bar{\Omega}^{2n}p(0, l_{\text{ssDNA}}, l_{\text{linker}})$.

Case 3: $\vec{l}_{\text{dsDNA}} \neq \vec{0}, \vec{l}_{\text{linker}} = \vec{0}$
We next turn to a diFab whose two ssDNA ends must end at the same point ($l_{\text{linker}} = 0$) but whose linker contains dsDNA ($\vec{l}_{\text{dsDNA}} \neq \vec{0}$) that can point in any direction. Analogous to the ssDNA segments, we assume the dsDNA has $\bar{\Omega}$ microstates (discretized by solid angle so that the length of the dsDNA segment does not affect this number).

As shown in Fig. S3D, the second random walk must end in a spherical shell of radius $l_{\text{dsDNA}}$ surrounding the endpoint of the first random walk (note that we neglect the negligible width of the DNA double helix). We combine the two random walks into a single random walk with $2\hat{n}$ steps by reversing the direction of the second random walk (pink) and translating it by $-\vec{l}_{\text{dsDNA}}$ so that it starts where the first random walk (purple) ends. Hence, the diFab can bind bivalently if and only if this effective random walk ends at a distance $l_{\text{dsDNA}}$ from the origin.

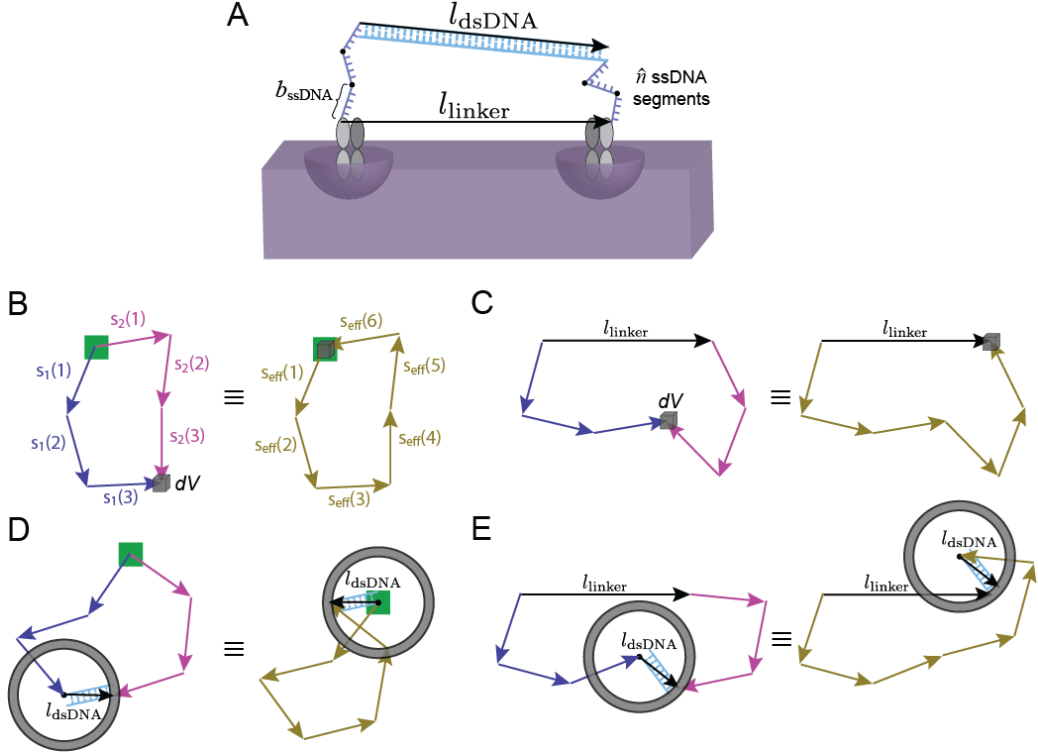

**Figure S3. The ideal chain model for a diFab linker.** (A) We compute the probability $p(l_{\text{dsDNA}}, l_{\text{ssDNA}}, l_{\text{linker}})$ that the ssDNA ($\hat{n}$ segments of length $b_{\text{ssDNA}}$) and dsDNA (1 rigid segment of length $l_{\text{dsDNA}}$) in the diFab linker will end at the appropriate distance $\vec{l}_{\text{linker}}$ required for bivalent binding. We consider the following cases: (B) $\vec{l}_{\text{dsDNA}} = \vec{0}, \vec{l}_{\text{linker}} = \vec{0}$; (C) $\vec{l}_{\text{dsDNA}} = \vec{0}, \vec{l}_{\text{linker}} \neq \vec{0}$; (D) $\vec{l}_{\text{dsDNA}} \neq \vec{0}, \vec{l}_{\text{linker}} = \vec{0}$; and the full model (E) $\vec{l}_{\text{dsDNA}} \neq \vec{0}, \vec{l}_{\text{linker}} \neq \vec{0}$. The two random walks start at each end of $\vec{l}_{\text{linker}}$, which is denoted by a green square when $\vec{l}_{\text{linker}} = \vec{0}$ (Panels B and D). Random walks end in the same location if they are within a small volume $dV$ of each other (Panels B and C) or if their ends are within $dV$ of the dsDNA of length $l_{\text{dsDNA}}$ (Panels D and E). In each case, the two ssDNA random walks are combined into a single effective random walk with $2\hat{n}$ steps.

Since $P(\vec{r})$ is radially symmetric about the origin, the number of microstates for any particular orientation of the dsDNA will be $\bar{\Omega}^{2\hat{n}} P(\vec{l}_{\text{dsDNA}}) dV$ and the total number of microstates considering all dsDNA orientations will be $\bar{\Omega}^{2\hat{n}+1} P(\vec{l}_{\text{dsDNA}}) dV$. However, in preparation for Case 4 below, it is instructive to compute the probability over linker states as the average over all dsDNA orientations,

$$p(l_{\text{dsDNA}}, l_{\text{ssDNA}}, 0) = \frac{dV}{4\pi l_{\text{dsDNA}}^2} \int_{\vec{x} \in B(\vec{0}, l_{\text{dsDNA}})} P(\vec{x}) d^2\vec{x}$$

$$= P(l_{\text{dsDNA}}) dV$$

$$= dV \left(\frac{3}{4\pi \hat{n} b_{\text{ssDNA}}^2}\right)^{\frac{3}{2}} e^{-\frac{3 l_{\text{dsDNA}}^2}{4\hat{n} b_{\text{ssDNA}}^2}}, \qquad \text{[S8]}$$

where $B(\vec{z}, r)$ represents the spherical shell of radius $r$ centered at $\vec{z}$, $P(\vec{x})$ is given by Eq. S5 with $\hat{n} \to 2\hat{n}$, and in the second step we used the radial symmetry of $P(\vec{x})$. The number of microstates is now given by $\bar{\Omega}^{2\hat{n}+1} p(l_{\text{dsDNA}}, l_{\text{ssDNA}}, 0)$ where the prefactor represents the $\bar{\Omega}^{2\hat{n}}$ orientations of the ssDNA and the $\bar{\Omega}$ orientations of the dsDNA. The similarity between Eqs. S7 and S8 reflect the symmetry between the dsDNA and the linker length in the system.

Case 4: $\vec{l}_{\text{dsDNA}} \neq \vec{0}, \vec{l}_{\text{linker}} \neq \vec{0}$

Finally, we turn to the case of a general diFab where the two ssDNA random walks are separated by a displacement $\vec{l}_{\text{dsDNA}}$ of dsDNA and must end with displacement $\vec{l}_{\text{linker}}$ from each other. As above, we transform these two random walks into a single effective random walk with $2\hat{n}$ steps that must finish in a spherical shell centered at $\vec{l}_{\text{linker}}$ with radius $l_{\text{dsDNA}}$ (Fig. S3E). Analogous to Case 3 above, the fraction of states of the dsDNA and ssDNA linker that allow bivalent binding is given by

$$p(l_{\text{dsDNA}}, l_{\text{ssDNA}}, l_{\text{linker}}) = \frac{dV}{4\pi l_{\text{dsDNA}}^2} \int_{\vec{x} \in B(\vec{l}_{\text{linker}}, l_{\text{dsDNA}})} P(\vec{x}) d^2\vec{x} \quad [S9]$$

where $B(\vec{z}, r)$ represents the spherical shell of radius $r$ centered at $\vec{z}$ and $P(\vec{x})$ is given by Eq. S5 with $\hat{n} \to 2\hat{n}$. This final integral is straightforward to evaluated analytically (see the Supplementary Mathematica notebook), yielding

$$p(l_{\text{dsDNA}}, l_{\text{ssDNA}}, l_{\text{linker}}) = \frac{dV}{l_{\text{linker}} l_{\text{dsDNA}}} \sqrt{\frac{3}{16\pi^3 \hat{n} b_{\text{ssDNA}}^2}} e^{-\frac{3(l_{\text{linker}}^2 + l_{\text{dsDNA}}^2)}{4\hat{n} b_{\text{ssDNA}}^2}} \sinh\left(\frac{3 l_{\text{linker}} l_{\text{dsDNA}}}{2\hat{n} b_{\text{ssDNA}}^2}\right). \quad [S10]$$

This leads to Eq. 2 with $b_{\text{ssDNA}} = 2\xi_{\text{ssDNA}}$, $\hat{n} = \frac{l_{\text{ssDNA}}}{2\xi_{\text{ssDNA}}}$, and with the constant $dV$ dropped (since it can be absorbed into $\alpha$). Therefore, the number of configurations for the bivalently bound diFab linker is given by $\Omega_2 = \bar{\Omega}^{2\hat{n}+1} p(l_{\text{dsDNA}}, l_{\text{ssDNA}}, l_{\text{linker}})$.

Note that $p(l_{\text{dsDNA}}, l_{\text{ssDNA}}, l_{\text{linker}})$ has no free parameters and is dictated purely by the geometry of each diFab. Furthermore, the factor $p(l_{\text{dsDNA}}, l_{\text{ssDNA}}, l_{\text{linker}})$ is the only term that varies between diFabs, whereas all remaining parameters (e.g., $K_D^{(1)}, E_2 - E_1$) are the same across all constructs. Hence, it is the solely the loss of entropy contained in $p(l_{\text{dsDNA}}, l_{\text{ssDNA}}, l_{\text{linker}})$ that determines how much better one diFab will be than another. Lastly, we point out that Eq. S10 is symmetric upon interchanging $l_{\text{dsDNA}}$ and $l_{\text{linker}}$, since every state of the ssDNA in a bivalently bound diFab with $l_{\text{dsDNA}}$ dsDNA and receptor binding sites spaced $l_{\text{linker}}$ apart would also enabled bivalent binding (with the locations of the dsDNA and the receptor interchanged) of a diFab with a length $l_{\text{linker}}$ of dsDNA binding to a receptor with binding sites spaced $l_{\text{dsDNA}}$ apart.

**The Number of Bivalent versus Monovalent Microstates for the Linker**

In this section, we now consider the flexibility $l_{\text{flex}}$ of the diFab and compute the full expressions for the number of microstates $\Omega_1$ and $\Omega_2$ of the monovalently bound and bivalently-bound Env spike shown in Fig. S3A. (We note that the number of microstates for the unbound state $\Omega_0$ need not be computed explicitly because the ratio of entropy and energy between the unbound and monovalently bound states are quantified by $K_D^{(1)}$ in Eq. 1.)

The number of microstates of the monovalently bound state is given by
$$\Omega_1 = \bar{\Omega}^{2\hat{n}+1}, \quad [S11]$$
where, as above, $\bar{\Omega}$ denotes the microstates of each segment in the linker (the $2\hat{n}$ ssDNA segments and the 1 dsDNA segment are all assumed to rotate freely in the ideal chain model). Note that this simple model neglects all interactions between the DNA, Fab, and Env including self-intersections.

When $l_{\text{flex}} = 0$, the multiplicity of the bivalently bound state was found above to be $\bar{\Omega}^{2\hat{n}+1} p(l_{\text{dsDNA}}, l_{\text{ssDNA}}, l_{\text{linker}})$ where $l_{\text{linker}}$ represents the distance spanned by the linker. If we approximate the direction of flexibility of the Fabs and the line joining the Fabs' C-termini to be colinear, the number of bivalent microstates is given by

$$\Omega_2 = \frac{\overline{\Omega}^{2\hat{n}+1}}{(\Delta l)^2} \int_{-\frac{l_{\text{flex}}}{2}}^{\frac{l_{\text{flex}}}{2}} \int_{-\frac{l_{\text{flex}}}{2}}^{\frac{l_{\text{flex}}}{2}} p(l_{\text{dsDNA}}, l_{\text{ssDNA}}, l_{\text{linker}} + x_2 - x_1) dx_2 dx_1$$

$$\approx \frac{\overline{\Omega}^{2\hat{n}+1} l_{\text{flex}}^2}{(\Delta l)^2} p(l_{\text{dsDNA}}, l_{\text{ssDNA}}, l_{\text{linker}}). \quad [\text{S12}]$$

In the first equality, $\Delta l$ represents a length scale that discretizes the flexibility imparted by the Fabs into the number of microstates. In the second equality, we assumed $l_{\text{flex}} \ll l_{\text{linker}}$ (since $l_{\text{flex}} \approx 20$ nm and $l_{\text{linker}} \approx 1$ nm) so that $l_{\text{linker}} + x_2 - x_1 \approx l_{\text{linker}}$ in the integrand. Substituting Eqs. S10 and S12 into Eq. S4, the relative probability of the bivalent and monovalent states takes the form

$$\frac{p_2}{p_1} = \frac{e^{-\beta(E_2-E_1)} dV}{(\Delta l)^2} \frac{l_{\text{flex}}^2}{l_{\text{linker}} l_{\text{dsDNA}}} \sqrt{\frac{3}{16\pi^3 \hat{n} b_{\text{ssDNA}}^2}} e^{-\frac{3(l_{\text{linker}}^2 + l_{\text{dsDNA}}^2)}{4\hat{n} b_{\text{ssDNA}}^2}} \sinh\left(\frac{3 l_{\text{linker}} l_{\text{dsDNA}}}{2\hat{n} b_{\text{ssDNA}}^2}\right)$$

$$\equiv \alpha n(l_{\text{dsDNA}}, l_{\text{ssDNA}}) \quad [\text{S13}]$$

where we have defined the prefactor $\alpha = \frac{e^{-\beta(E_2-E_1)} dV}{(\Delta l)^2}$ containing the unknown constants that are independent of the diFab and Env and $n(l_{\text{dsDNA}}, l_{\text{ssDNA}})$ to be the remaining geometry-dependent terms as per Eq. 2. Using the normalization condition $p_0 + p_1 + p_2 = 1$, we find that the % neutralization of HIV-1 for a diFab linked together by dsDNA and ssDNA is given by

$$p_1 + p_2 = \frac{\frac{[\text{Ab}]}{K_D^{(1)}} + \frac{[\text{Ab}]}{K_D^{(1)}} \alpha n(l_{\text{dsDNA}}, l_{\text{ssDNA}})}{1 + \frac{[\text{Ab}]}{K_D^{(1)}} + \frac{[\text{Ab}]}{K_D^{(1)}} \alpha n(l_{\text{dsDNA}}, l_{\text{ssDNA}})}. \quad [\text{S14}]$$

As a final aside, we note that the ideal chain model is valid when there are at least 3 segments in the random walk (Fig. S4), which requires $3 b_{\text{ssDNA}} \approx 14$ bases of ssDNA in the diFab. Constructs whose ssDNA has at least this many bases should be well characterized by our model, whereas diFabs with less ssDNA would require the more complicated worm-like chain model.

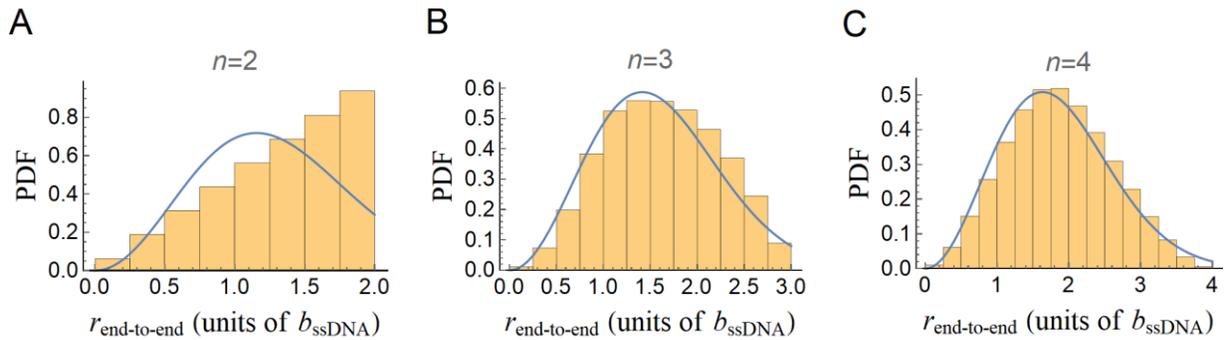

**Figure S4. The ideal chain model applies for ≥ 3 segments.** Comparing the analytic form of the ideal chain model in the large $\hat{n}$ limit (Eq. S5) with numerical simulations of a 3D random walk with (A) $\hat{n} = 2$, (B) $\hat{n} = 3$, and (C) $\hat{n} = 4$ segments. The y-axis shows the probability density function of the trials. As few as three segments are required to closely match the large $\hat{n}$ limit.

## C. Generalizing the Model to GappedFabs and TriFabs

In this section, we discuss how the model of diFab neutralization (Eq. 4) can be generalized to account for either gappedFabs with ssDNA breaks within the dsDNA segment or triFabs that combine three Fabs to achieve greater avidity effects.

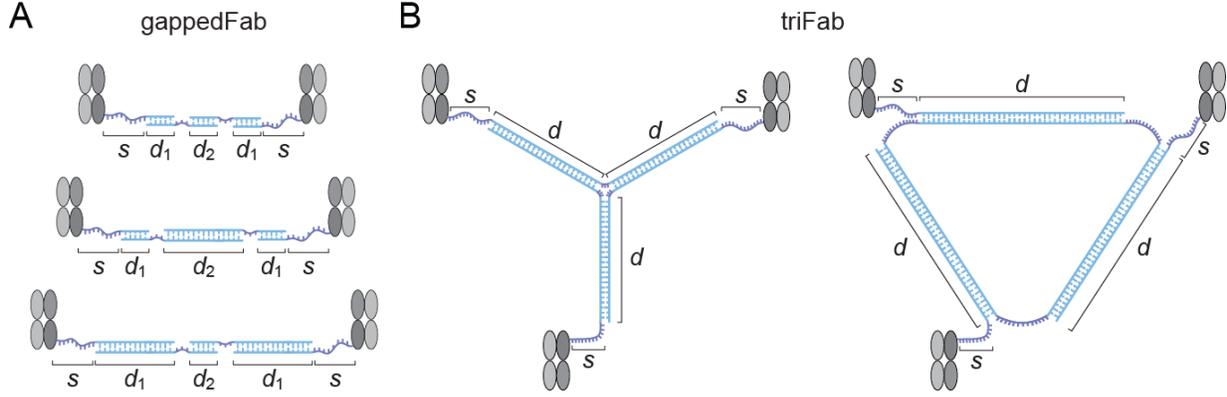

**Figure S5. Generalizing the model to other DNA-Fab designs.** (A) GappedFabs have ssDNA break in their dsDNA, providing flexibility while keeping each segment rigid. (B) Two alternative triFabs designs (different from the triFab design in Fig. 1B) composed of three Fabs connected by DNA.

### GappedFabs represent a Random Walk of dsDNA

Here, we consider the effects of placing small ssDNA breaks within the dsDNA as shown in Fig. S5A. For simplicity, we restrict our analysis to linkers composed of a middle segment with $d_2$ bp dsDNA, surrounded by $n_1$ segments of $d_1$ bp dsDNA (each connected by short ssDNA segments; $n_1 = 2$ in the three gappedFabs shown in Fig. S5A), and flanked by $s$ bases ssDNA on either side. This structure enables us to consider both the two outer ssDNA segments and the $n_1$ dsDNA segments as random walks about the single dsDNA segment of length $d_2$ (and it is straightforward to generalize to arbitrary gappedFab designs). In doing so, we assume the short ssDNA segments connecting the dsDNA act as free hinges with negligible lengths.

Eq. S9 in Appendix B shows that in the case $n_1 = 0$ where the linker must span the distance $\vec{l}_{\text{linker}}$, the combined ssDNA random walk ($2n$ segments of length $b_{\text{ssDNA}}$) starting at the origin must end on a sphere of radius $l_{\text{dsDNA}}$ around $\vec{l}_{\text{linker}}$. When $n_1 > 0$, the combined ssDNA random walk ends at the arbitrary point $\vec{z}$, and the dsDNA random walk must then start at $\vec{z}$ and end on a sphere of radius $l_{d_2} = d_2(0.34\frac{\text{nm}}{\text{bp}})$ around $\vec{l}_{\text{linker}}$, representing the length of the $d_2$ segment. Using the probability density Eq. S5 for a random walk, the probability that the linker in a gappedFab will bivalently bind is given by

$$p_{\text{gap}}(l_{\text{dsDNA}}, l_{\text{ssDNA}}, l_{\text{linker}}) = \int_{\vec{x} \in B(\vec{l}_{\text{linker}}, l_{d_2})} \int_{\vec{z} \in \mathbb{R}^3} P_{\text{ssDNA}}(\vec{z}) P_{\text{dsDNA}}(\vec{x} - \vec{z}) d^3\vec{z}\, d^3\vec{x} \qquad [S15]$$

where

$$P_{\text{ssDNA}}(\vec{z}) = \left(\frac{3}{2\pi b_{\text{ssDNA}}^2(2n)}\right)^{\frac{3}{2}} e^{-\frac{3z^2}{2b_{\text{ssDNA}}^2(2n)}} \qquad [S16]$$

represents the probability that the endpoint of the ssDNA random walk starting at the origin will end at $\vec{z}$,

$$P_{\text{dsDNA}}(\vec{x} - \vec{z}) = \left(\frac{3}{2\pi l_{d_1}^2 n_1}\right)^{\frac{3}{2}} e^{-\frac{3|\vec{x}-\vec{z}|^2}{2 l_{d_1}^2 n_1}} \quad [\text{S17}]$$

denotes the probability that the endpoint of the dsDNA random walk starting at $\vec{z}$ will end at $\vec{x}$, and $l_{d_1} = d_1 (0.34 \frac{\text{nm}}{\text{bp}})$. Replacing $p \to p_{\text{gap}}$ and $\overline{\Omega}^{2n} \to \overline{\Omega}^{2n+n_1}$ in Eq. S12 yields the desired number of microstates for the bivalent configuration.

Fig. S6 shows how the potency of the previously constructed diFab ($d$=62, $s$=12) gets worse (larger IC$_{50}$) as the dsDNA segment is broken up into $\hat{n}$ pieces of equal length ($n_1 = \hat{n} - 1$, $d_1 = d_2 = \frac{62}{\hat{n}}$, $s = 12$). Since the $\hat{n}$ segments of dsDNA will be much more confined when the diFab is bivalently bound relative to the monovalent configuration, the entropic penalty of bivalent binding quickly increases with the number of segments leading to lower potency.

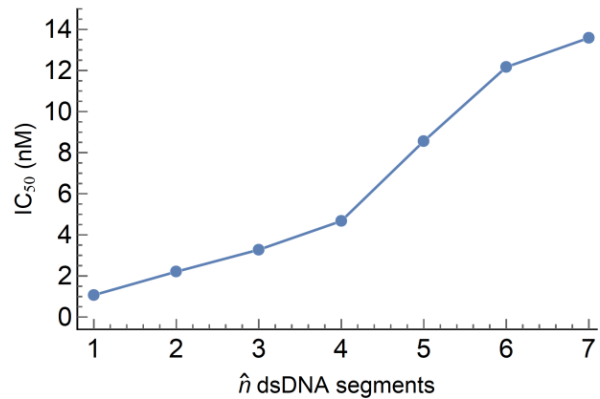

**Figure S6. Potency of a gapped diFab made up of 62 bp dsDNA broken into $\hat{n}$ equal length pieces.** Breaking up the dsDNA increases the entropic cost of bivalent binding, decreasing its potency.

### TriFabs exhibit Greater Avidity than diFabs

Previously, we showed that the diFab with ($d$=62, $s$=12) has an IC$_{50}$ that is 100x smaller than its one-armed Fab counterpart (21). This raises the question of whether a construct with additional Fabs could further reduce the IC$_{50}$. Here, we outline how our model can be extended to consider a linear triFab (shown in Fig. 1B), and this calculation can be readily extended to the alternate triFab designs in Fig. S5B. The calculation below predicts that the optimal triFab will be 100x more potent than the optimal diFab, providing a method to leverage the knowledge of the HIV-1 Envelope spike we derived from our synthetic diFabs to engineer even more potent reagents.

As in Appendix B, we model the ssDNA segments with $s$ bases as random walks and neglect both the self-intersection of these random walks as well as intersections with the Env spike, though we note that these effects may be more prominent in a triFab than in a diFab. Furthermore, we neglect the combinatorics characterizing which binding arm attaches to an Env binding site (e.g., there are $\binom{3}{1}$ ways to bind monovalently; $\binom{3}{2}$ ways to bind bivalently, although in the linear triFab the simultaneous binding of nearest neighbor arms will be different from the binding of the two arms furthest apart; and $\binom{3}{3}$ ways to bind trivalently) as well as the details of these configurations (e.g., there are six possible configurations of binding the three triFab arms to the three Env epitopes).

With these assumptions, the Boltzmann statistical weights for the triFab (analogous to those in Fig. 2A for the diFab) are $1$, $\frac{[Ab]}{K_D^{(1)}}$, $\frac{[Ab]}{K_D^{(1)}} \alpha\, n$, and $\frac{[Ab]}{K_D^{(1)}} \alpha^2 n^2$ for the states with 0, 1, 2, and 3 Fab arms bound, respectively. Analogous to Eq. S14, the probability of binding (and hence neutralizing) a spike is given by

$$p_1 + p_2 + p_3 = \frac{\frac{[Ab]}{K_D^{(1)}} + \frac{[Ab]}{K_D^{(1)}} \alpha\, n + \frac{[Ab]}{K_D^{(1)}} \alpha^2 n^2}{1 + \frac{[Ab]}{K_D^{(1)}} + \frac{[Ab]}{K_D^{(1)}} \alpha\, n + \frac{[Ab]}{K_D^{(1)}} \alpha^2 n^2} \qquad [S18]$$

with an IC$_{50}$ given by

$$\text{IC}_{50}^{\text{triFab}} = \frac{K_D^{(1)}}{1 + \alpha\, n + \alpha^2 n^2}. \qquad [S19]$$

As described in the main text, this implies that the boost in avidity in going from an optimal diFab to an optimal triFab is equal to the boost in going from a Fab to an optimal diFab.